\documentclass[a4paper,aps,prl,showpacs,twocolumn,superscriptaddress]{revtex4}
\usepackage{graphicx,t1enc}
\begin{document}
\date{\today}
\title{Spatial features of population dynamics arising from  \\mutual
interaction of different age groups in rodents}

\author{M. N. Kuperman}
\affiliation{Consortium of the Americas for Interdisciplinary Science and
         Department of Physics and Astronomy, University of New Mexico, Albuquerque,
         New Mexico 87131, USA}

\affiliation{Centro At{\'o}mico
Bariloche and Instituto Balseiro, 8400 S. C. de Bariloche,
Argentina\\
Consejo Nacional para las Investigaciones
Cient\'{\i}ficas y T\'ecnicas, Argentina}
\author{V. M. Kenkre }
\affiliation{Consortium of the Americas for Interdisciplinary Science and
         Department of Physics and Astronomy, University of New Mexico, Albuquerque,
         New Mexico 87131, USA}

\vspace{1cm}

\begin{abstract}
\ \\
We study the dynamics of the transmission of the hanta virus infection among mouse populations, taking into account, simultaneously,  seasonal variations of the
environment and interactions within two classes in the mouse population: adults and subadults. The interactions considered are not symmetric between the two age-organized classes and are responsible for driving the younger members away from home ranges. We consider the case of a bounded habitat  affected by seasonal variations.
\ \\
\end{abstract}


\maketitle

\section{Introduction}
Theoretical investigations of population dynamics of animals such as mice or other rodents derive their current importance both from their direct relevance to the spread of epidemics such as the Hantavirus \cite{YATES,AK,PASI,KPHA,KLAC} but also from the interplay of the approaches of physics and ecology that such investigations encourage \cite{MURR,LI,COSNER,Levin,KOT}. Our interest in the present paper is on prevalent effects of the competitive interaction among adults and
subadults among rodent populations. Several ecological studies   have shown the importance of such an interaction \cite{fgt1,fgt2}.
The interaction can be studied  by considering an age-structured population of mice, i.e., by splitting the population into classes organized by age.
There is competitive struggle for territory  associated with
the fact that the occurrence of infected cases of several types
of Hantavirus is mostly observed among adult individuals. This, in
turn, is related to the proposed mechanism of infection among
rodents: territorial fights that produce wounds on the animals as a tangible and verifiable consequence. The
subadult population, consisting of smaller-sized individuals, tends to maintain its distance
from the fights, and is thus less susceptible to infection. The effect of remaining not too close to the source of the fights, viz., the adult population,  subadult individuals are forced to abandon
the already colonized spaces and are therefore driven to new habitats, most
of the time less suitable for survival.

We propose here a model considering an age-structured population
composed by the two well differentiated groups we have mentioned above: subadults and adults. A
third group, composed by juvenile individuals may be taken into
account but will be left out of our calculations as relatively unimportant.

Like previous models \cite{AK,PASI,KPHA,KLAC}, ours is based on a set of Fisher-like equations \cite{FISHER} to describe the evolution of the
population of  individuals of each group whose densities we will respectively denote by $M_a$ and $M_y$. Here, the suffix $a$ stands for adults and the suffix $y$ for the young individuals, the subadults.

\section{The model}

The specific set of equations in our model would be, in the absence of territorial interactions,
\begin{eqnarray}
\frac{\partial M_a}{\partial t}&=&D_a\nabla^2 M_a + \mu M_y -\delta M_a- \frac{M_a M}{K(t)},
\nonumber \\
\label{fish} && \\
\frac{\partial M_y}{\partial t}&=&D_y\nabla^2 M_y +\beta M_a - \mu M_y - \frac{M_y
M}{K(t)}\nonumber.
\end{eqnarray}
Here  we are considering different diffusion coefficients $D_a$ and
$D_y$ for the adults and the subadults, to take into account the fact that adult individuals tend to
remain in a rather bounded area, considered as each individual
territory, while the subadults tend to move further and probably faster. One would therefore expect $D_a<D_y.$ The
term in each of the above equations which is associated with the competition for resources, is characterized by what is
called the environmental parameter $K$ and is proportional to the so-called carrying capacity.  We will take $K$ to be time
dependent because of seasonal variations. We denote the total population of the mice by $M=M_a+M_y$.
The rate $\mu$ is one of  transformation of subadults into adults through the process of maturity, thus is proportional to the subadult population and appears with reversed sign in the two equations. The subadults are born from the adults at  $\beta$, the birth
rate. Death of the adults occurs at rate
$\delta$ and we have omitted it from the subadult population equation for simplicity and to reflect the simplifying assumption that natural death visits only the adults. If predators were introduced into our considerations different rates might be put into the equations including a death rate for the subadults.

In this work we propose to modify the above set of equations to take into account the ecologically important experimental observation
related to territorial fights. Due to competition for the conquest and preservation of the home range, adult mice
tend to fight among themselves. Indeed, these fights have been suspected to constitute one of the most important ways of transmission of the Hantavirus among rodents of the same species. As a result of these territorial threats of the adults, the subadults, being smaller in size, are forced to abandon already colonized environments and move towards unoccupied spaces.
There are at least two different ways to include this tendency into the equation. To take into account the fact that subadults will tend to move away from places with high adult occupation, we may assume the subadult flux to  be proportional to the
gradient of the adult population, $\nabla M_a$; at the same time, no subadult flux is possible in the absence of young individuals. Thus, the interaction term should be proportional to the population of the subadults $M_y$ as well as to the adult gradient. The resulting set of equations   is
\begin{eqnarray}
\frac{dM_a}{dt}&=&D_a\nabla^2 M_a + \mu M_y -\delta M_a- \frac{M_a M}{K(t)}
\nonumber \\
\label{chem1} && \\
\frac{dM_y}{dt}&=&D_y\nabla^2 M_y +\beta M_a - \mu M_y - \frac{M_y
M}{K(t)} \nonumber \\
&&+\kappa\nabla(M_y\nabla M_a) \nonumber
\end{eqnarray}
with $\kappa$ a constant that determines the strength of the interaction.

The other manner that young individuals might tend to get away from populated areas is by creating a flux of their own class which is proportional to the local density of the other class, i.e., the  adult. Under these conditions we will have
\begin{eqnarray}
\frac{dM_a}{dt}&=&D_a\nabla^2 M_a + \mu M_y -\delta M_a- \frac{M_a M}{K(t)}
\nonumber \\
\label{chem2} && \\
\frac{dM_y}{dt}&=&D_y\nabla^2 M_y +\beta M_a - \mu M_y - \frac{M_y
M}{K(t)} \nonumber \\
&&+\kappa\nabla(M_a\nabla M_y). \nonumber
\end{eqnarray}
Comparison of the two new interaction expressions,
$$\nabla(M_y\nabla M_a)=(\nabla M_a)(\nabla M_y)+M_y\nabla^2 M_a$$
and
$$\nabla(M_a\nabla M_y)=(\nabla M_a)(\nabla M_y)+M_a\nabla^2 M_y$$
shows that they share a common term proportional to the gradient of each of the classes and another term which in one case is proportional to
the density of subadult population and the Laplacian of the adult population and in the other case the situation is precisely reversed.

In both cases there is a homogeneous  nontrivial non-negative steady state solution, for Eqs. (\ref{fish}), (\ref{chem1}) or (\ref{chem2})
that can be written in terms of the parameters of the problem  $\beta$, $\delta$, $\mu$ and $K$, and is independent of the interaction strength $\kappa$:

\begin{eqnarray}
M_{a}^0&=&\frac{K}{2( \beta - \delta ) }(\delta (\delta+A) + \mu (2\beta-\mu+A)) \nonumber \\
\nonumber && \\
M_{y}^0&=& \frac{K}{2( \beta - \delta ) }[(\mu
 ( \delta -\mu +{A})\nonumber
 \\
  &&         - \beta( \delta + 3\mu -
          {A} )) ]
 \label{sol1}
\end{eqnarray}
where $A =\sqrt{\delta^2 + 4\beta \mu - 2\,\delta \mu + \mu^2}.$
In the following section we will study  Eqs.(\ref{chem1})and (\ref{chem2}), i.e. we will analyze the effects of each of the new terms separately.

We are interested in effects of temporal or seasonal changes in the environment. These changes   will be reflected in a change in the carrying capacity
of the habitat. Therefore, we will consider a varying $K$.
The ability of a given species to adapt to a changing environment is related to their mobility as well as to other quantities such as the birth and death rates.
One way to characterize the mobility is by considering the so-called Fisher velocity of the species associated with traveling environments which can naturally arise as the seasons change. When favorable conditions move in space and time, it is important to find out how the  population can or cannot follow them, and whether there are critical velocities of the traveling environment which separate parameters regions in which the species survive or undergo extinction.   To facilitate the subsequent discussion we introduce the concept of a refugium, a bounded domain with a high carrying capacity. The population of a given species can live within this region.
Outside the refugium, the living conditions are too harsh for the species to survive.

\section{Static refugium}

\begin{figure}
\includegraphics[width=9cm]{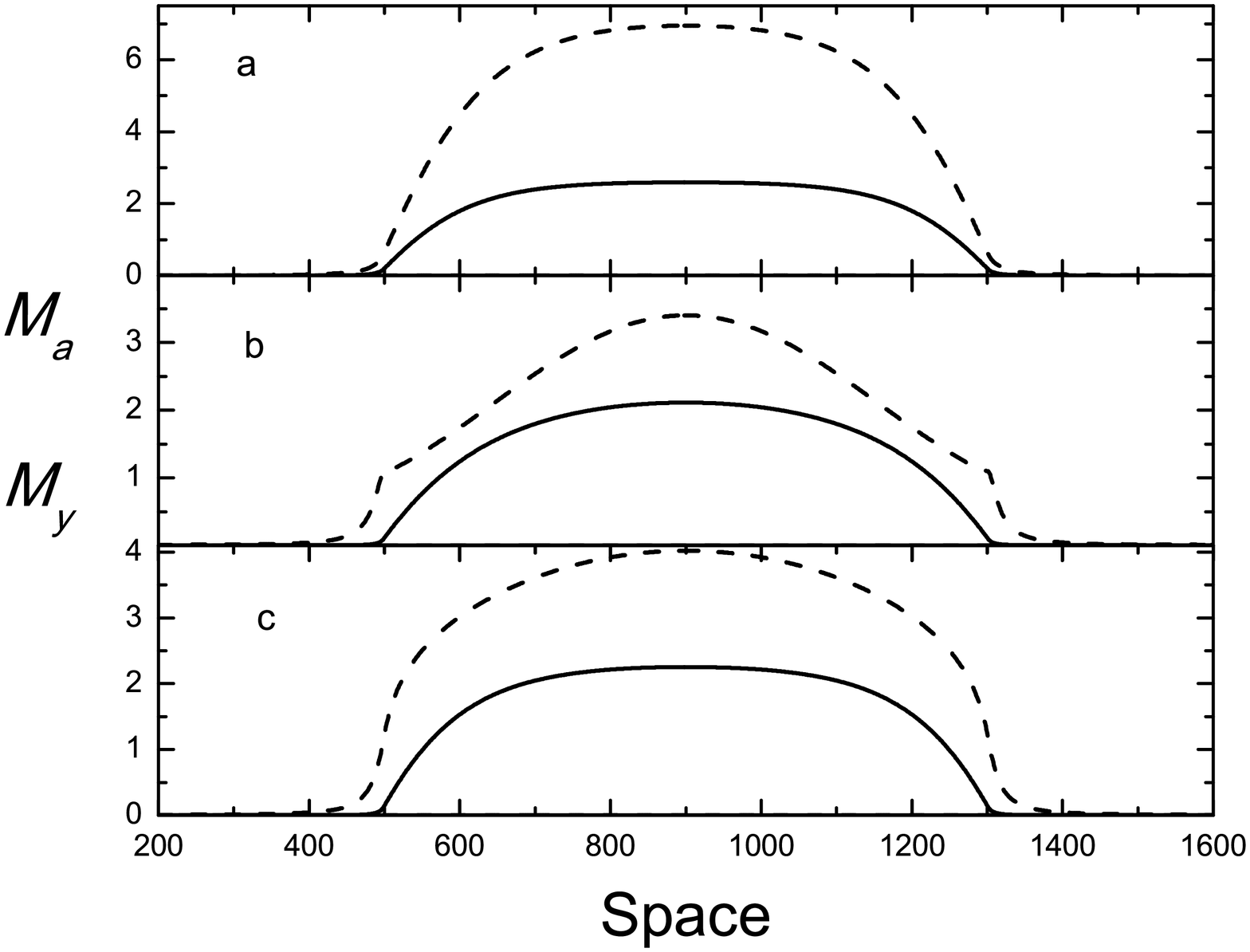}
\caption{Steady state profiles of $M_a$ (solid)and $M_y$ (dashed)in a bounded domain. The plots correspond to the solution to (a) Eqs.(\ref{fish}), (b) Eqs.(\ref{chem1}), (c) Eqs.(\ref{chem2}). Case (a) represents no interaction whereas cases (b) and (c) describe the two different interactions considered (see text).  Parameters (arbitrary units) are $\beta=4$, $\mu=0.4$, $\delta=0$, $\kappa=10$. Additionally the environment parameter $K$ is $10^{-4}$ (practically zero) outside and 9 inside the refugium.}
\label{front1}
\end{figure}

When  no temporal variations of the refugium are considered, we find  stationary profiles for both populations which differ from those previously found when considering Fisher-like coupled equations.
In Figs.\ref{front1} we display the profiles of both populations in three cases. All parameters of the problem are the same in all cases, with the exception of $\kappa$ which is $0$ in (a) but, in arbitrary units, $10$ in (b) and (c).

We can see that while the profile of the adult density $M_a$ remains almost unchanged in the presence of interactions, that of the subadult density $M_y$ presents interesting effects.
Since the interactions tend to drive the subadults away from the adults, in both cases there is an increase of the number of individuals at the borders of the refugium. The effect of the interaction term is more evident in regions where the gradient of $M_a$ is greater. This fact explains the shoulder on the profile of $M_y$ in Fig. \ref{front1}.b. On the other side, the profile in Fig. \ref{front1}.c presents the smoothest shape,   consistent with a greater effective diffusion coefficient.

\begin{figure}
\includegraphics[width=9cm]{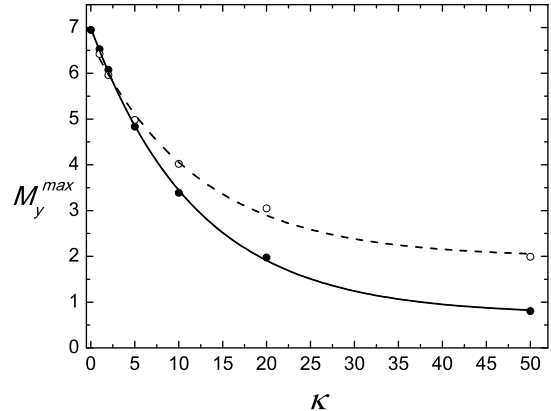}
\caption{Maximum value of the subadult mouse population $M_y$ as a function of the strength $\kappa$ of the interaction between subadults and adults from Eqs. (2)  (full circle, solid line ) and Eqs. (3) (empty circle, dashed line). The curves are the result of exponential fitting of the solutions obtained numerically.}
\label{inic}
\end{figure}

The interaction term affects not only the shape of the profile of both population but also the maximum values attained by the population inside the refugium. We note that the solutions in Eq.(\ref{sol1}), where $\kappa$   plays no role, correspond to infinite domains. When bounded domains are considered,  $D_a$, $D_y$, as well as $\kappa$ and the domain size, affect the maximum value of the  population. We plotted In Fig. \ref{inic} the behavior of the maximum attained by $M_y$, $M_y^{max}$, for a given size of the refugium and constant diffusion coefficients as a function of the values of $\kappa$.
The population of subadults undergoes an overall decay  as $\kappa$  increases. The curves in Fig. (\ref{inic}) were fitted with  decaying exponentials. A mono-exponential fit works in each case.

An interesting effect is the possibility of maintaining a subadult population \emph{localized} when it is bounded by a population of adults as if they were forming a well confining the younger mice. As the interaction terms
prevent the population of subadults from moving towards an increasing gradient of  adults, the subadult population  remains
localized until the population of adults has decreased sufficiently (by diffusion) to reach a state in which diffusion and
interaction in the subadult population can compete and delocalize the nucleus of subadults.

\section{traveling refugium}

\begin{figure}
\includegraphics[width=9cm]{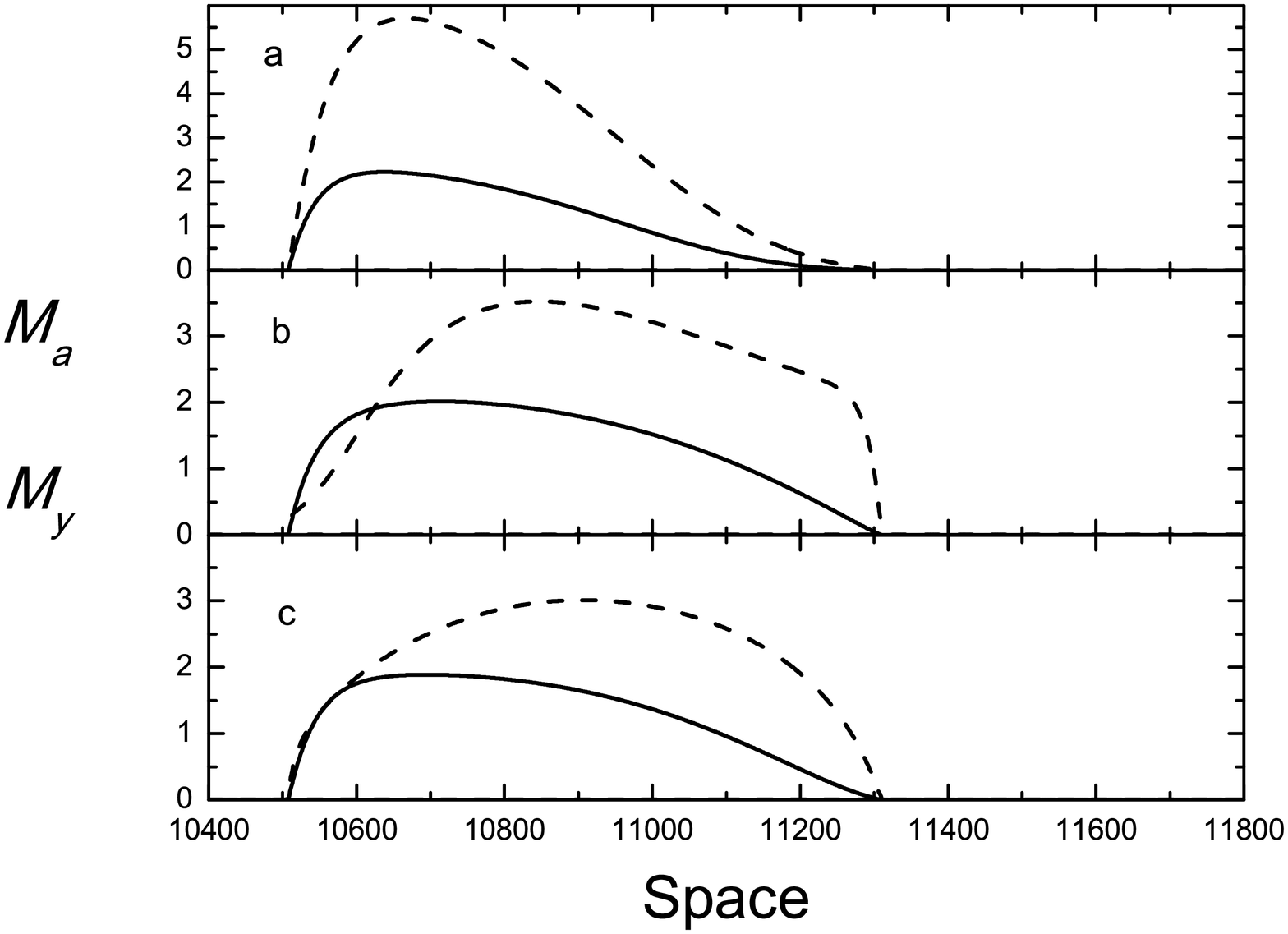}
\caption{Steady state profiles of $M_a$ (solid)and $M_y$
(dashed)in a traveling bounded domain with velocity $v=1.3$. The
plots correspond to the solution to (a) Eqs.(\ref{fish}), (b)
Eqs.(\ref{chem1}), (c) Eqs.(\ref{chem2}). With $\beta=4$,
$\mu=0.4$, $\delta=0$, $\kappa=10$ and $K=9$ inside the bubble.
Parameters are in arbitrary units.} \label{frontv}
\end{figure}

Consider now a  traveling refugium of constant size. This will
allow us to test the ability of the population to survive while
following the moving environment. We will find the critical
velocity of the refugium $v_c$, above which survival is no longer
possible. Fig. \ref{frontv} displays  the changes in the fronts as
the refugium starts to move in each of the three cases.

Again, while the profile of $M_a$ (adults) remains almost unchanged, the profile of $M_y$ (subadults) shows some effects. Both interactions
(expressed in Eqs. \ref{chem1} and \ref{chem2} tend to maintain  the symmetry of the distribution of the population despite the translational
movement;  the effect of Eqs.  \ref{chem2} is with stronger intensity as it strongly biases the population contrary to the movement.

When considering only one species and the Fisher equation, the critical velocity is given by the Fisher velocity $v_f=\sqrt{Da}$,
with $D$ the diffusion constant and $a$ the birth rate. When dealing with a set of coupled equations it is still possible to show
that there is a critical velocity, associated closely with the critical velocity of the slowest species.

If the new terms in Eqs.(\ref{chem1}) and (\ref{chem2}) are neglected to get Eqs. (\ref{fish}),  within the range of the values
used for the parameter values of both populations, adult and subadult densities can be shown to be of the same order throughout the whole domain.
The adult population would then be described by
\begin{equation}
\frac{dM_a}{dt}=D_a\nabla^2 M_a + \mu M_a -\delta M_a- \frac{M_a^2}{K}.
\label{sola}
\end{equation}
\begin{figure}
\includegraphics[width=9cm]{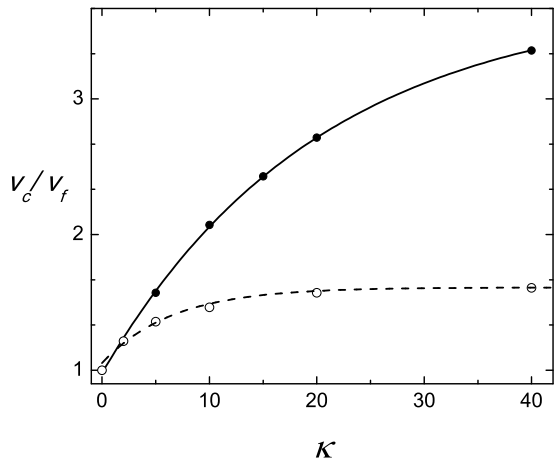}
\caption{Critical velocity of the population (relative to the
Fisher velocity) as a function of $\kappa$ from Eqs. (2) (full
circle, solid line ) and Eqs. (3) (empty circle, dashed line). The
curves are exponential fitting of the data.} \label{vcrit}
\end{figure}
This would mean that we can associate a Fisher velocity,
$v_f=\sqrt{D_a(\mu-\delta)}$ with the population $M_a$. While it
appears that a similar argument might be used for $M_y$, this is
not true because the idea that both populations are of the same
order is no longer valid in that case when the velocity of the
refugium is above the Fisher velocity for $M_a$. The adults are
slower than subadults. It is therefore that the \emph{lower} Fisher
velocity appropriately describes the critical situation. This
explains the numerical results that show that when considering
Eqs(\ref{fish}), the critical velocity of the entire population
$v_c$ is almost the same  as the critical velocity that can be
obtained from Eq.(\ref{sola}).

We display in Fig. \ref{vcrit} interaction effects on the critical
velocity $v_{c}$ by plotting $v_{c}/v_{f}$ (where $v_{f}$ is the
Fisher velocity of the adult mice in the absence of interactions)
as a function of the interaction strength $\kappa$ for the two
kinds of interaction. We find that the numerical solutions may be
fitted excellently by saturating exponentials. We observe that in
both cases there is an increase in  $v_c$ as $\kappa$ increases.
The effect is much more evident for Eq.{\ref{chem1}} as was
expected from the shape presented by its profile in Fig.
\ref{frontv}b.

\begin{figure}
\includegraphics[width=9cm]{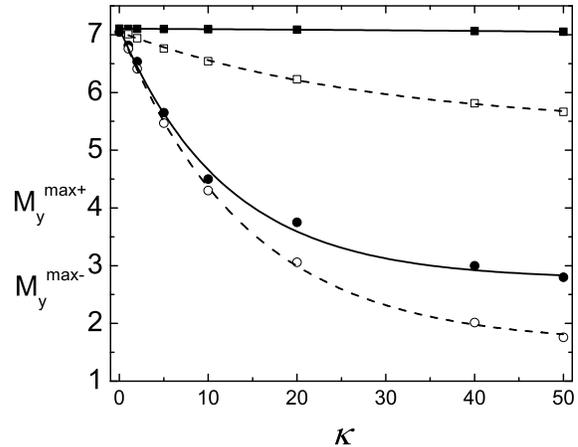}
\caption{Limiting maximum values $M_y^{max+}$(squares) and
$M_y^{max-}$(circles). Full symbols correspond to Eqs. (2) while
empty symbols correspond to Eqs. (3). Curves are exponential
fittings of the data} \label{ampli}
\end{figure}

\section{Breathing refugium}
\begin{figure}
\includegraphics[width=9cm]{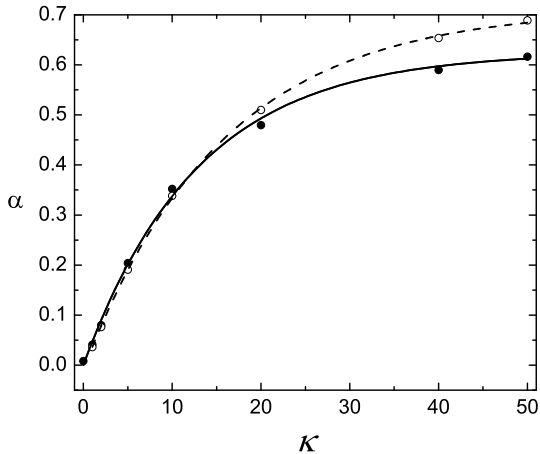}
\caption{Relative amplitude of the temporal oscillations of the maximum value of $M_y(x,t)$ as a function of $\kappa$ from Eqs. (2 )(full circle,
solid line ) and Eqs. (3) (empty circle, dashed line). The curves are exponential fitting of the data. }
\label{oscili}
\end{figure}

Seasonal variations  of the relevant parameters may correspond to
the refugium being  centered in a static point with its size
changing in an oscillating way. We call this case a breathing
refugium. To show the new features due to the effect of the
interaction terms, we plot the maximum of the subadult density
$M_y$ as a function of time. If there are no interactions between
subadults and adults, the maximum suffers negligible periodic
variations. On the other hand, in the presence of interactions,
the maximum  oscillates between two values displayed in Fig.
\ref{ampli} by squares and circles, respectively. We call these
two limiting maximum values $M_y^{max+}$ and  $M_y^{max-}$. The
amplitude of this variation increases in a very apparent way as
$\kappa$ grows. This is shown in Fig. \ref{oscili}, where we plot
the relative difference
$\alpha=\frac{M_y^{max+}-M_y^{max-}}{M_y^{max+}}$, which is the
ratio of the difference between the limiting maximum values to the
greater of these maximum values.

\begin{figure}
\includegraphics[width=9cm]{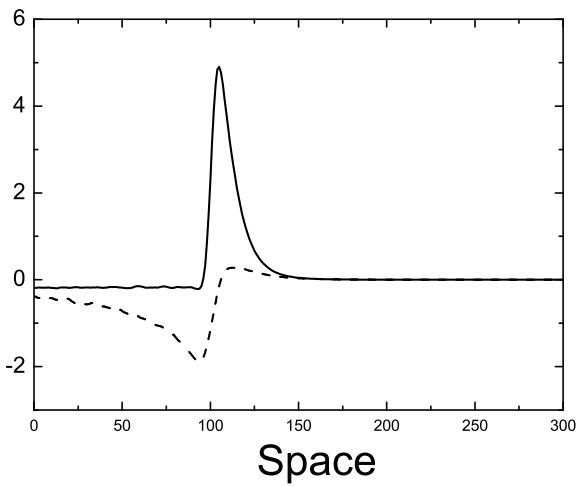}[h]
\caption{Comparative intensity of each effect on the border of the
moving front of subadult population. The effect associated to Eqs.
(\ref{chem1}) is in solid line, while the one corresponding to
Eqs. (\ref{chem2}) is in dashed line. The units are arbitrary}
\label{compa}
\end{figure}

\section{Conclusions}
We have shown the emergence of some interesting effects due to the
inclusion of interaction among different groups in a population.
We have also compared our new results with those already known from simple Fisher equations with no interactions. Among the
new findings are the change in the shape of the
steady and traveling profiles (Figs. \ref{front1} and
\ref{frontv}), the increase of the critical velocity of the
populations (Fig. \ref{vcrit}) with interaction strength, and the oscillatory behavior of the
population profiles when seasonal changes in the environment are
considered (Figs. \ref{ampli} and \ref{oscili}). Some of these
results can be understood easily  in the context of Fig. \ref{frontv}. When a
traveling refugium is considered, the steady fronts displayed in
Fig. \ref{front1} change. The traveling profile are asymmetric,
showing a depletion of the concentration of individuals at the
head (right) of the front. In the presence of the interaction, the
subadult population tends to move towards areas less populated
with adults. This explains the changes in the traveling profiles
presented by both population in Figs. \ref{frontv}a and
\ref{frontv}b. The effect of the term in Eqs. \ref{chem1}  is stronger than that of the one considered in  Eqs.\ref{chem2}.
Indeed, we observe that not only the population of subadults moves
to the right in a more evident way but the change in the critical
velocity is also more intense. 

The shape presented by the profiles associated to each of
the equations can be understood by qualitative arguments. As mentioned before, a comparison of
the two new interaction expressions, shows that they share a
common term proportional to the gradient of each of the classes
$\nabla(M_a)(\nabla M_y)$. The other terms, 
$M_y\nabla^2 M_a$
and
$M_a\nabla^2 M_y$ respectively,   promote a flux of the subadult
population towards less populated regions, i. e. to the right of
the traveling front. This explains the shift of the profile of
subadults. The difference in the resulting profile can
be understood when analyzing the difference in intensity of ech
term, as shown in Fig. \ref{compa}, where we compare the intensity
of each of the effect when considering a stationary profile. The
oscillations found when the refugium is breathing are due to the
fact that a system described by any of the  Eqs. (\ref{chem1}) or
(\ref{chem2}) is much more sensitive to changes in the size of the
refugium, as can be observed by the values displayed in Fig.
\ref{inic}, where without loss of generality, only one size of the
refugium was analyzed.

This work was supported in part by the NSF under grant no.
INT-0336343 and by NSF/NIH Ecology of Infectious Diseases under
grant no. EF-0326757.

\end{document}